\documentclass[review,10pt]{elsarticle}
\usepackage{lineno,hyperref}
\modulolinenumbers[1]
\journal{}
\biboptions{authoryear}

\usepackage{pgfplots}
\usepackage{float}

\usepackage{booktabs}
\usepackage{multirow}

\usepackage{mathptmx}

\DeclareMathAlphabet{\mathcal}{OMS}{cmsy}{m}{n} 

\usepackage{amsmath} 

\usepackage{amstext}
\usepackage{amssymb}
\usepackage{esint}
\usepackage{siunitx}

\usepackage[color=red!40]{todonotes}

\usepackage{geometry}
\usepackage{graphicx}

\makeatletter
\@ifundefined{showcaptionsetup}{}{%
	\PassOptionsToPackage{caption=false}{subfig}}
	
\usepackage[font=normalsize]{subfig}

\makeatother

\newcommand{\quotes}[1]{``#1''}

\begin{document}
\begin{frontmatter}

\title{Operator-based machine learning framework for generalizable prediction of unsteady treatment dynamics in stormwater infrastructure}

\author[address1]{Mohamed Shatarah}
\author[address1]{Kai Liu}
\author[address1]{Haochen Li \corref{mycorrespondingauthor}}
\cortext[mycorrespondingauthor]{Corresponding author}
\ead{hli111@utk.edu}
\address[address1]{Water Infrastructure Laboratory, Department of Civil and Environmental Engineering, University of Tennessee, Knoxville, Tennessee 37996, USA}

\begin{abstract}
Stormwater infrastructures are decentralized urban water management solutions that often experience uncontrolled and unsteady hydraulic and pollutant loadings driven by episodic rainfall-runoff events. Evaluating their in-situ treatment dynamics and performance remains a fundamental challenge for cost-effective system design, implementation, and planning. To date, lumped dynamic models, such as the continuously stirred tank reactor (CSTR) model, have been employed to account for these unsteady loads. While computationally efficient, these models oversimplify physical reality, limiting their predictive capability and insights into treatment dynamics and design optimization. Computational fluid dynamics (CFD) offers robust predictive capabilities by resolving the underlying turbulent transport and pollutant fate physics; however, its high computational cost constrains its application for simulating stormwater infrastructure dynamics under unsteady loading conditions, particularly over extended periods. To address this limitation, this study leverages state-of-the-art operator learning and develops a composite neural network (CPNN) framework to predict the spatial and temporal dynamics of stormwater infrastructure under unsteady hydraulic and particulate matter (PM) loading. This framework is applied and evaluated on a common urban stormwater treatment system, a hydrodynamic separator (HS). The results demonstrate a high predictive capability of the proposed CPNN framework, achieving an $R^2$ score above 0.8 in 95.2\% of cases for hydrodynamic predictions. For PM concentration predictions, the $R^2$ score exceeds 0.8 in 72.6\% of cases and remains between 0.4 and 0.8 in 22.6\%. The study also highlights challenges in learning spatial-temporal dynamics under complex loading conditions spanning multiple orders of magnitude. Specifically, the model exhibits weaker performance under extreme low-flow conditions due to their reduced contribution to the total training loss function. Furthermore, this study explores the unique capability of CPNN for automatic differentiation, assessing the sensitivity and influence of storm event loading conditions on PM transport dynamics in the HS. The potential of CPNN for evaluating stormwater infrastructure performance over continuous and long-term periods is discussed. This work represents a critical step toward enabling robust and climate-aware stormwater infrastructure planning and implementation.

\end{abstract}

\end{frontmatter}


\section{Introduction} \label{sec:Introduction}
Stormwater infrastructure plays a vital role in urban water management by controlling runoff, mitigating flooding, and reducing pollutants such as suspended solids, nutrients, and metals in surface water \citep{Okaikue-Woodi2020}. The importance and benefits of stormwater infrastructure are widely recognized. In the United States (US), the bipartisan infrastructure law, passed in 2021, allocates more than 50 billion USD to the Environmental Protection Agency (EPA) to enhance drinking water, stormwater, and wastewater infrastructure \citep{infraBill2023}. Among these funds, 11.7 billion USD is designated for the Clean Water State Revolving Fund (CWSRF), which directly supports stormwater projects. In Europe, the European Investment Bank (EIB) granted approximately 1 billion euros to support the management of wastewater, stormwater, and solid waste between 2019 and 2023 \citep{EIB2024}. Despite continuous investments and collaborative efforts, a fundamental understanding of in-situ system dynamics and the ability to evaluate transient treatment performance prior to implementation remain significant challenges. These knowledge gaps can contribute to substantial uncertainty in assessing and quantifying the overall benefits of stormwater infrastructure, particularly when longer evaluation periods are considered. This poses a significant obstacle to the development of science-based regulatory guidance for cost-effective planning and implementation of stormwater infrastructures.

One of the primary challenges in examining the in situ treatment performance of stormwater systems lies in the inherent unsteadiness of hydraulic conditions and pollutant loads \citep{Spelman2018}. Unlike unit processes and operations in water and wastewater treatment facilities, where system loads are generally controlled and regulated based on plant capacity, stormwater systems are decentralized and implemented across urban watersheds. These systems are exposed to highly variable and uncontrolled hydraulic and pollutant loads. In-situ field monitoring has revealed that hydraulic loads passing through stormwater systems during a single rainfall event can vary by as much as 100 times, depending on watershed characteristics and rainfall patterns. Similarly, pollutant loads can fluctuate drastically over time as the storm progresses, with higher concentrations typically observed during the early stages of an event (often referred to as the \quotes{first flush effect}) \citep{Sansalone2005, Lee2000}. Under such unsteady conditions, system hydrodynamics and the coupled pollutant transport exhibit complex, nonlinear behavior, resulting in time-dependent pollutant removal efficiency \citep{Wilson2009}. Indeed, studies have confirmed the importance of accounting for these unsteady loading conditions and their underlying transport dynamics when evaluating the in-situ treatment performance of stormwater systems \citep{Sansalone2009, Pathapati2011, Garofalo2018}.

Recognizing the dynamic nature of stormwater systems, researchers have moved beyond the current state-of-practice and regulations, which are primarily based on conceptual quasi-steady-state approximations and lumped metrics, such as Surface Overflow Rate (SOR) and Event Mean Concentration (EMC) \citep{Liu2019}. Various methods have been developed to predict treatment performance under unsteady loads. A classic approach is the CSTR model, also known as a lumped dynamical model, originally proposed in the 1950s \citep{Reynolds1996}. In the CSTR model, which assumes a well-mixed condition ($\nabla(\cdot) = 0$), momentum transport is neglected, and the mass transport equations are simplified into ordinary differential equations (ODEs), with unsteady loads incorporated through source terms. The simplicity of solving ODE systems has made the CSTR model widely used in controlled process modeling where the well-mixed condition is valid \citep{Huber2006, Li2022a}. However, for stormwater treatment systems, where mass transport is strongly coupled with momentum exchange driven by episodic rainfall events, neglecting momentum balance and turbulent mixing in the CSTR model can lead to a distorted representation of physical processes and less accurate predictions of system treatment performance \citep{Garofalo2020, Li2021basin}. With advancements in numerical methods and computational hardware, CFD has emerged as a powerful tool for simulating water infrastructure \citep{Liu2011, Ying2012, Mendoza2016, Liu2020, Yee2023, Cata2023}. Unlike the CSTR model, where momentum transport is ignored, CFD directly solves the coupled momentum and mass transport equations, thereby preserving the spatial and temporal variations inherent in stormwater dynamics. Studies have demonstrated that CFD provides substantially improved predictive capabilities compared to the lumped CSTR approach \citep{Dickenson2009, Garofalo2020, Li2021c, Li2023a}.

The robust predictive capabilities offered by CFD come at a significantly increased computational cost. For instance, a single Reynolds-Averaged Navier-Stokes (RANS) CFD simulation of a 125-minute storm event can require up to 120 hours of computation on a Dell Precision 690 machine equipped with two quad-core Intel Xeon 2.33 GHz processors and 16 GB of RAM \citep{Garofalo2011}. Furthermore, each CFD simulation is tailored to specific loading conditions, meaning that different load scenarios require entirely separate simulations. This leads to a rapid escalation in computational costs, making CFD impractical to evaluate a stormwater treatment system under various loading conditions or over longer implementation periods. To mitigate computational costs and enable robust evaluation of stormwater treatment performance, researchers have explored approximating unsteady treatment dynamics using a series of steady-state CFD simulations. Methods such as Stepwise Steady Flow \citep{Pathapati2011}, Instantaneous Stepwise Steady (ISS) \citep{Cho2013}, and Implicit Solution Stepwise Steady (IS3) \citep{Spelman2017} have been developed with this goal in mind. These pioneering approaches show promise in reducing computational requirements but are fundamentally constrained by their reliance on the stepwise steady-state assumption. Specifically, this assumption requires that the system's flow relaxation timescale be significantly smaller than the timescale over which inflow varies. For stormwater systems experiencing a wide range of loading conditions, this assumption is rarely valid, especially at lower flow rates or in systems designed with higher residence times (i.e., larger flow relaxation timescales) \citep{Li2021basin}. These limitations underscore the need for innovative approaches that can address the computational constraints while robustly reconstructing the transient dynamics of unsteady stormwater systems.

Emerging fields such as artificial intelligence (AI) and machine learning (ML) present promising solutions to overcome these challenges. ML models have been adopted across various water systems for predicting water quality and optimizing treatment processes \citep{Sun2019, Mullapudi2020, Huang2021, Garzon2022, Zhang2025}. Artificial neural networks (ANNs) have demonstrated remarkable capabilities in nonlinear mapping within high-dimensional spaces and have been widely applied in fields such as computer vision, natural language processing, and fluid dynamics \citep{Lowe2022, Li2022b}. In the area of stormwater infrastructure, research has shown that ML can effectively learn complex nonlinear relationships, such as those between clarification basin loading conditions, geometric parameters, and treatment performance \citep{Li2022c, Li2022d, Li2023b, Li2024a}, achieving coefficients of determination ($R^2$) above 0.8 for 66.4\% of cases and an $R^2$ above 0.4 for 89.2\% of cases. Recent advancements in operator-based ML methods, including DeepONet, MIONet, and Fourier Neural Operator (FNO) \citep{Li2020c, Lu2021, Wang2021, Jin2022, Li2024b} allow ML models to learn parameterized inputs. This capability enables the model to efficiently generalize across a wide range of scenarios and configurations, rather than being constrained to specific cases. Building on these advancements, this study develops and evaluates an operator-based ML framework for learning and reconstructing the flow hydrodynamics and particulate matter transport under unsteady loading conditions. Unlike studies that focus on lumped treatment performance metrics, this research aims to simulate and resolve the unsteady three-dimensional spatial and temporal dynamics of flow hydrodynamics and PM transport within stormwater infrastructure under a variety of loading scenarios. This is a complex and underexplored challenge but represents a critical step toward the ultimate goal of enabling robust and efficient evaluations of stormwater infrastructure performance under diverse climatic conditions and extended evaluation periods \citep{Spelman2018a}. This framework also extends our prior work on operator learning for stormwater systems \citep{Li2024a}, progressing from quasi-steady inputs to unsteady loading conditions and from a simplified two-dimensional (2D) system model to a full three-dimensional (3D) model. The specific objectives of this study are: (1) to develop a robust CPNN operator-based ML framework for unsteady flow and transport dynamics, (2) to examine the predictive capability of the framework across a range of hydraulic and PM loading conditions, and (3) to leverage the CPNN's automatic differentiation capability for evaluating the 3D unsteady HS system sensitivity to storm event loading conditions.

\section{Methodology}

As an outline, Section \ref{formulation} describes the parameterization of hydraulic and PM loading conditions and associated operator learning tasks. Section \ref{operator} presents the formulation and architecture of the developed CPNN framework. Section \ref{training} details the CPNN model training strategy and evaluation metrics.

\subsection{Formulation of hydrodynamics and PM transport prediction as an operator learning task} \label{formulation}

\begin{figure}[H]
\centering
\includegraphics[width=1.0\textwidth]{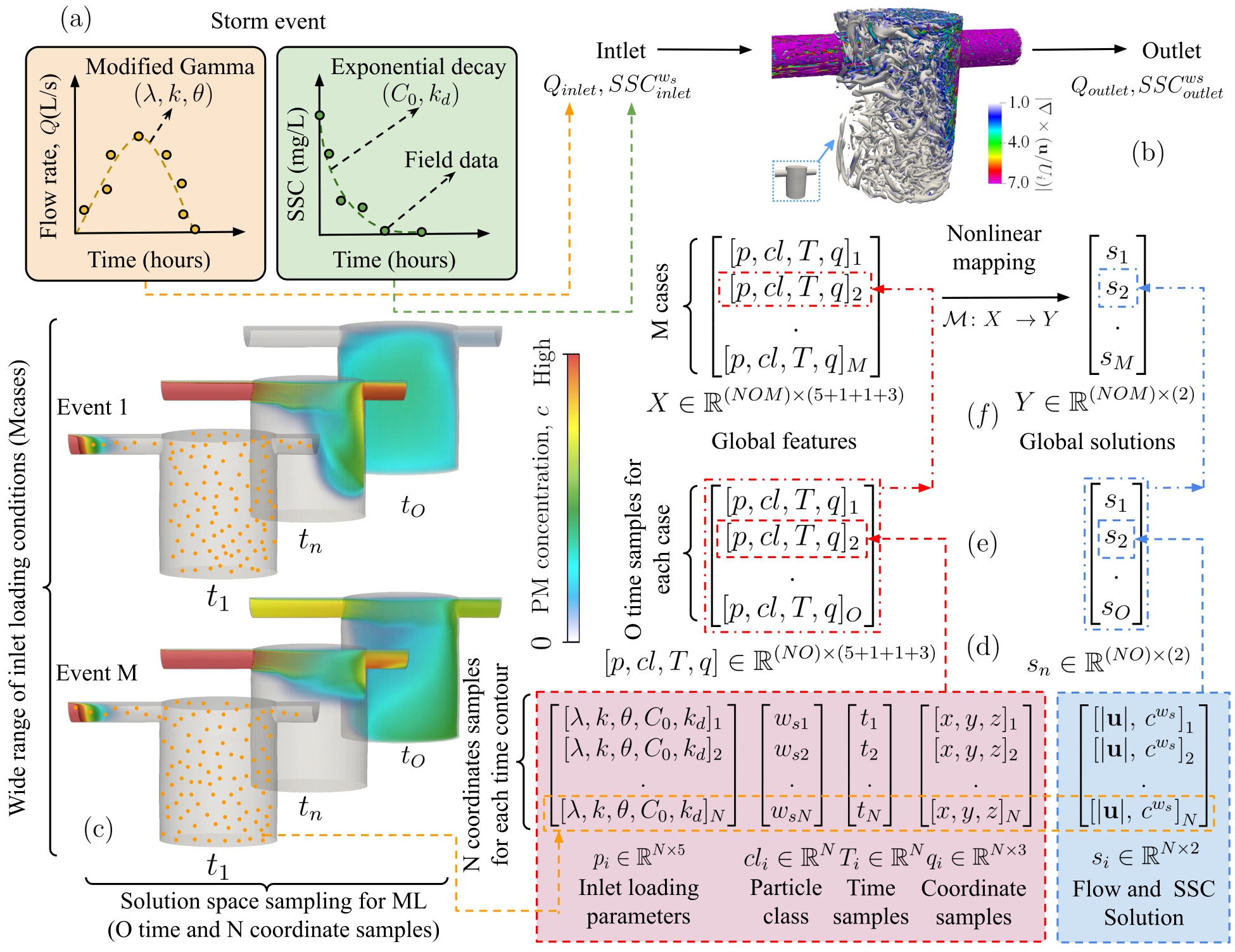}
\caption{Formulation of stormwater system hydrodynamics and PM fate prediction as an operator learning task. (a) Parameterization of the inlet loadings hydrograph as a modified gamma function and the pollutograph as an exponential decay function, (b) Visualization of hydrodynamic separator (HS) system. As the influent containing suspended sediment (SSC) enters the tank, a portion of the particles settles, whereas the remainder leaves with the effluent through the outlet, (c) Examples of CFD simulations of PM transport in the HS treatment dynamics for a range of $M$ loading conditions, $O$ time steps and $N$ spatial samples, (d) Data layout for $N$ spatial samples at each time instance of each scenario of loading conditions, (e) Data layout for $O$ time instances of each storm scenario, and (f) Global data layouts for the entire $M$ storm events. $k$, $\theta$, and $\lambda$ are the hydrograph modified gamma distribution function parameters. $C_0$ and $k_d$ are the pollutograph exponential decay function parameters. $w_s$ is the PM terminal velocity, which is used to identify each class. $t$ is the sampling point's time stamp. $x$, $y$, $z$ are the spatial coordinates for the sampling point. $|\mathbf{u}|$ and $c$ are velocity magnitude and PM concentration. $\mathbb{R}$ denotes the matrix dimensions.}
\label{fig:formulation}
\end{figure}

Fig. \ref{fig:formulation} encapsulates the formulation of turbulent transport and PM fate in stormwater infrastructure as an operator-learning task. A common stormwater treatment system, the hydrodynamic separator, is considered to examine and demonstrate the efficacy of the proposed formulation and architecture. The formulation presented is generalizable and can be adapted to other water infrastructure. The HS is selected here because (1) the HS is widely deployed throughout urban drainage systems \citep{EPAStormwater2021} and (2) the CFD solver used in this study was extensively benchmarked and validated under controlled laboratory experiments for a large number of commercially available HS systems \citep{Li2020a,Li2020b,Li2021b}. The specific HS geometry considered in this study follows the full-scale system tested by \citet{Howard2010thesis,Li2024c}. It has a diameter and a depth of 1.21 m. The inlet and outlet are configured along the centerline. The HS functions as a clarification system. During a storm event, runoff carrying PM enters the HS through the inlet. Under the influence of gravitational force, PM begins to diverge from the flow path. A portion of the loaded PM is separated from the carrying stream and retained in the system. The overall PM removal is the result of the nonlinear interaction between turbulence and PM transport \citep{Li2021a}.

As shown in Fig. \ref{fig:formulation}a, the inlet flow hydrograph and PM pollutograph (measured as suspended sediment concentration, SSC) are parameterized by a modified gamma function and an exponential decay function as defined in Eqs. \ref{eq:gamma_function}-\ref{eq:exponential_function}. Table \ref{tab:parameter_summary} summarizes the parameter ranges for these two functions, which are determined based on regression analysis of field observed data for 40 runoff events at an urban watershed between June 2010 and September 2016 \citep{Dickenson2012, Spelman2017}, as shown in Figs. S1-S4 in the online Supplemental Materials. This approach of using parameterized functions to represent the hydrograph and pollutograph is adopted to reduce the parameter space in the subsequent ML model training and enhances the model's generalizability.

\begin{equation}
    Q(t)={\frac{\lambda}{\gamma (k) {\theta}^{k}}}t^{k-1}e^{-t/\theta},\label{eq:gamma_function}
\end{equation}
\begin{equation}
    C(t) = C_0 \exp\left(-k_d t\right). 
    \label{eq:exponential_function}
\end{equation}
In these equations, $k$ is the shape parameter, $\theta$ is the scale parameter, and $\lambda$ is a scaling factor. $C_0$ is the initial concentration parameter,  $k_d$ is the decay concentration coefficient.

\begin{table}[H]
\caption{Parameter ranges for the parameterized hydrograph and PM pollutograph. ($\lambda$, $k$, and $\theta$ are the scaling factor, the shape parameter, and the scale parameter. $C_0$ and $k_d$ are initial concentration and decay coefficient. These parameter ranges are determined based on field-observed data.)}
\centering
\setlength\tabcolsep{4pt} 
\begin{tabular}{cccccc}
\toprule 
Parameter & $\lambda$ & $k$ & $\theta$ & $C_0$ & $k_d$ \\
\midrule
Lower bounds & 0.0017 & 1.1 & 0.23 & 0.1072 & 0.5\\
Upper bounds & 0.2012 & 99.3 & 51.5 & 3.6963 & 1.0\\
\bottomrule
\end{tabular}
\label{tab:parameter_summary}
\end{table}

Based on this parameterization, a series of higher-fidelity CFD simulations is performed for turbulent transport and PM fate in the HS system. The CFD model applied here is based on the unsteady Reynolds-Averaged Navier-Stokes (URANS) equations and implemented through an open-source C++ CFD library, OpenFOAM. The CFD model uses the Euler-Euler method to simulate the PM transport and considers one-way coupling between PM and the flow hydrodynamics. This one-way coupling is considered because the flow is primarily driven by the pressure gradient induced by the rainfall runoff and, to a lesser extent, by PM-induced stratification (i.e., density difference) \citep{Li2020a}. In the simulation, 9 PM classes are used to discretize the particle size distribution (PSD), which are represented by terminal velocities $w_s$ ranging from \SI{1e-6}{m/s} to \SI{0.1}{m/s}. The corresponding diameters of these PM terminal velocities are provided in Table S1 in the online Supplemental Materials. Fig. \ref{fig:formulation}b visualizes the hydrodynamic separator system considered in this study. Influent containing SSC enters the HS, where some PM settles while the remaining suspended solids exit with the effluent through the outlet. As illustrated by Fig. \ref{fig:formulation}c, simulations are conducted for a range of storm events and loading conditions. Each simulation case is subjected to a distinct hydrograph and pollutograph generated through Latin hypercube sampling (LHS) over the identified parameter ranges (Table \ref{tab:parameter_summary}). A one-hour storm duration is considered in the simulation. These simulations are performed using the ISAAC-NG high-performance computing (HPC) facility at the University of Tennessee. Fig. \ref{fig:formulation}c shows examples of volumetric renderings of PM concentration at three time instances under two different loading conditions.

In total, 640 ($M$) unsteady 3D CFD simulation cases  are carried out, as shown in Fig. \ref{fig:formulation}f. Each case contains 360 ($O$) time instances that are sampled at 10-second intervals, as shown in Fig. \ref{fig:formulation}e. Additionally, 8,000 ($N$) spatial samples are generated for each case through LHS. As illustrated in Fig. \ref{fig:formulation}d, each sampling point has the following set of features: the loading parameters $\mathbf{p}_i = \left[ \lambda, k, \theta, C_0, k_d \right]$, the PM class $\mathbf{cl}_i = [w_s]$, the time points $T_i = [t]$, and the spatial coordinates $ \mathbf{q}_i = \left[x, y, z \right]$. The corresponding solution is $s_i= [|\mathbf{u}|, c^{w_s}]$. The global feature matrix $\mathbb{X}$ encompasses $M$ cases, $O$ time instants, $N$ coordinate points, leading to a combined dimension of $(NMO, 5+1+1+3)$. The global solution matrix $\mathbb{Y}$ has a dimension of $(NMO,2)$. With this formulation, the prediction of HS treatment dynamics under unsteady loading conditions is transformed into a supervised operator-based ML task. The goal is to learn the nonlinear operator $\mathcal{M}: \mathbb{X} \rightarrow \mathbb{Y}$ that maps global feature matrix $\mathbb{X}$ to the global solution matrix $\mathbb{Y}$. 

\subsection{Operator-based machine learning architecture for time-dependent system}
\label{operator}
\begin{figure}[H]
\centering
\includegraphics[width=0.9\textwidth]{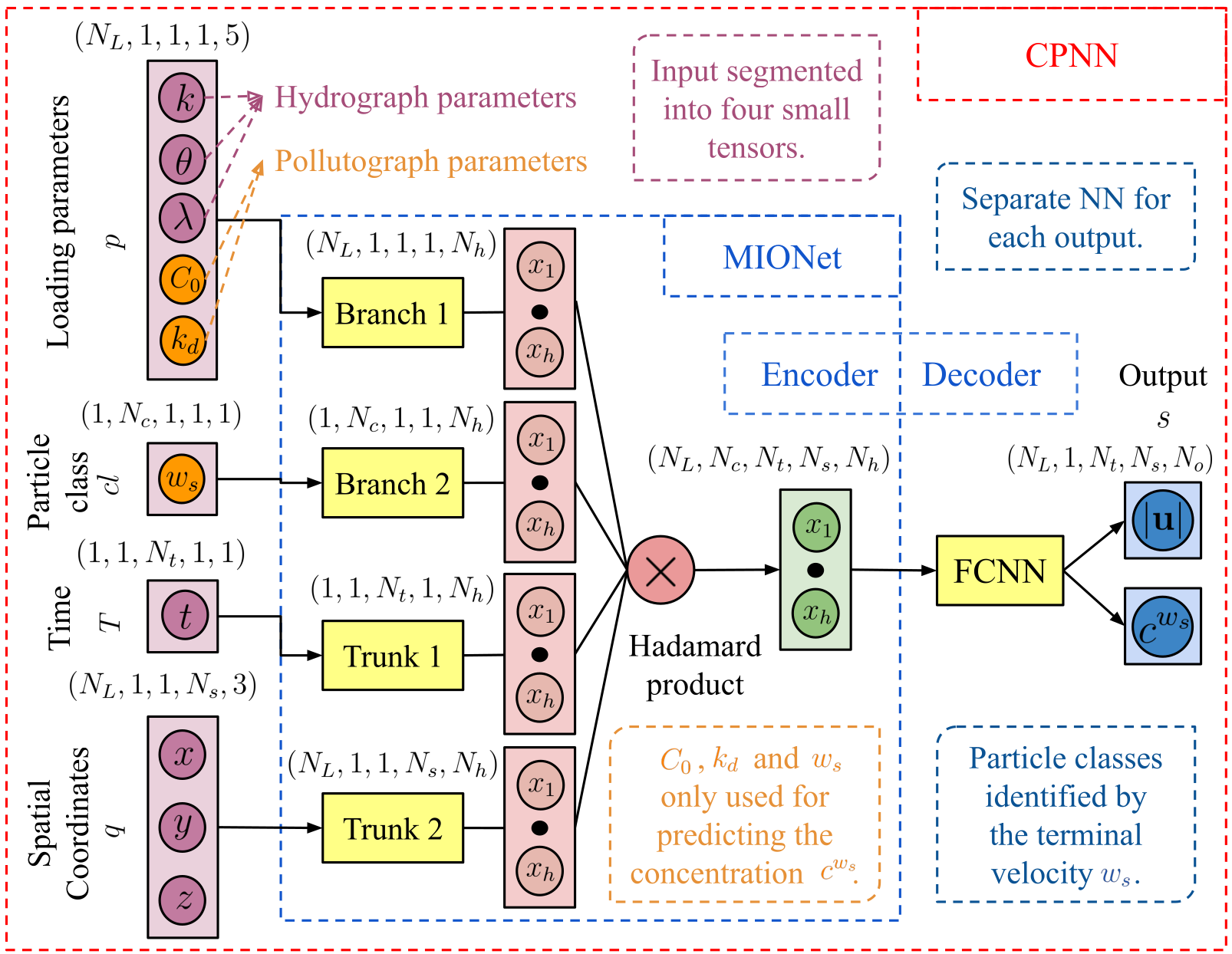}
\caption{Visualization of the composite neural network (CPNN) architecture. MIONet serves as an efficient feature encoder for stormwater system event inputs, which include loading parameters $\mathbf{p}$, particle class $\mathbf{cl}$, time stamps $\mathbf{T}$, and continuous coordinates $\mathbf{q}$. The architecture comprises two trunks and two branches integrated through element-wise multiplications. This is followed by a fully connected neural network (FCNN) to better process the system's nonlinear dynamics. Finally, the CPNN predicts the desired system output solution $\mathbf{s}$.}
\label{fig:architecture}
\end{figure}

Various methods have been developed for learning the nonlinear operator $\mathcal{M}$. Building upon the foundation of the universal approximation theorem for operators \citep{Chen1995}, the deep operator network (DeepONet) was developed to learn the nonlinear operators that map between function spaces \citep{Lu2021, Mao2021}. The subsequent work by \citet{Jin2022} extends the DeepONet to multiple input functions and develops a multiple-input operators network. As illustrated in Fig. \ref{fig:architecture}a, MIONet can incorporate multiple branch and trunk networks. Two trunk networks and two branch networks are used for this application. This architecture is particularly useful for scenarios where a system's behavior is influenced by multiple independent input groups. In the context of a stormwater treatment system, the branch networks (Br1 and Br2) process event loading parameters ($\mathbf{p}$) and PM class ($\mathbf{cl}$). Simultaneously, the trunk networks (Tr1 and Tr2) process temporal information ($\mathbf{T}$) and the spatial coordinates ($\mathbf{q}$). These neural network (NN) layers effectively encode the relevant information from each input group into separate latent representations.

A key benefit of MIONet is its capacity to capture the nonlinear interactions among these multiple input features while minimizing memory usage. This architecture has a significant advantage over standard ANN and DeepONet architectures \citep{Lu2021, Li2024a}. Unlike DeepONet, which primarily couples two input functions, MIONet can embed a series of input features through element-wise multiplication between multiple branch and trunk networks, as defined in Eq. \ref{eq:MIONet}. This approach allows MIONet to capture the dependencies between independent features such as event loading conditions, PM class, time stamp, and 3D spatial coordinate within the stormwater system. MIONet can achieve significant input size reduction compared to both standard ANN and DeepONet. For example, as shown in Eq. \ref{eq:ANN}, the standard ANN repeats the spatial coordinates for each time instance, then repeats both the spatial coordinates and time stamp for each PM class, and finally repeats all these features for each loading case. This results in a larger input of size $\mathbf{p} \times \mathbf{cl} \times \mathbf{T} \times \mathbf{q}$, containing a significant amount of redundant data. In contrast, the MIONet breaks the input into the smallest independent input groups (four in this study), then leverages feature embedding through the Hadamard product (point-wise multiplication) between the four separate inputs, eliminating redundant data and resulting in an input of size $\mathbf{p} + \mathbf{cl} + \mathbf{T} + \mathbf{q}$. This significantly reduces the input size (by 100x in this study) compared to ANN while maintaining comparable predictive capability.
\begin{equation}
\mathrm{ANN}(\mathbf{p}, \mathbf{cl}, \mathbf{T}, \mathbf{q}) = \mathbf{NN}(\mathbf{p} \times \mathbf{cl} \times \mathbf{T} \times \mathbf{q}),
\label{eq:ANN}
\end{equation}
\begin{equation}
\mathrm{DeepONet}(\mathbf{p}, \mathbf{cl}, \mathbf{T}, \mathbf{q}) = \mathbf{Br}(\mathbf{p} \times \mathbf{cl} \times \mathbf{T}) \odot \mathbf{Tr}(\mathbf{q}),
\label{eq:DeepONet}
\end{equation}
\begin{equation}
\mathrm{MIONet}(\mathbf{p}, \mathbf{cl}, \mathbf{T}, \mathbf{q}) = \mathbf{Br1}(\mathbf{p}) \odot \mathbf{Br2}(\mathbf{cl}) \odot \mathbf{Tr1}(\mathbf{T}) \odot \mathbf{Tr2}(\mathbf{q}).
\label{eq:MIONet}
\end{equation}
In these equations, $\times$ denotes a Cartesian product of the inputs before they are processed by the NN. This forms all possible pairs of elements from the involved input sets (e.g., pairing each element from one set with every element of another). By contrast, $\odot$ represents an element-wise Hadamard product of the latent feature vectors at the end of DeepONet and MIONet architectures. This production merges the outputs of separate branch or trunk networks.
 
Leveraging the efficient parameter encoding capabilities of MIONet, a CPNN is developed for efficiently predicting the unsteady three-dimensional evolution of hydrodynamics and particulate matter transport in stormwater infrastructure. As shown in Fig. \ref{fig:architecture}, the CPNN is formed by combining feature encoding through MIONet with feature decoding through a fully connected neural network (FCNN). This formulation shares a similar spirit with position encoding techniques in natural language processing (NLP) \citep{Vaswani2017, Tancik2020} and builds upon our previous research \citep{Li2024a} for predicting basin hydrodynamics and PM transport. In the CPNN, the MIONet encodes the input loading parameters, PM classes (i.e., terminal velocity), time stamps, and spatial coordinates into a latent feature, which is then decoded by an FCNN through a series of hidden layers, as defined in Eq. \ref{eq:CPNN}. The hyperbolic tangent function (Tanh) is used as the activation function between layers in this study. Additionally, the Leaky ReLU activation function is also investigated, as shown in Fig. S6 in the online Supplemental Material. This architecture combines efficient parameter encoding capabilities of MIONet for multiple trunks and branches and enhanced expressivity in spatial and temporal dynamics of FCNN, as shown in the work of \citep{Li2024a}. Note that in this study, the CPNN framework does not involve physics-informed training. A study by \citet{Krishnapriyan2021} shows that the partial differential equation (PDE) constraint in physics-informed neural network (PINNs) can become highly ill-conditioned, leading to slow training.
\begin{equation}
\mathbf{y}(\mathbf{p}, \mathbf{cl}, \mathbf{T}, \mathbf{q};\boldsymbol{\Theta}) = \mathbf{W}_m \big \{\mathbf{H}_{m-1} \circ \cdots \circ \mathbf{H}_1 \circ \mathbf{M} \big \} (\mathbf{p}, \mathbf{cl}, \mathbf{T}, \mathbf{q}) + \mathbf{b}_m, \,\, 
\mathrm{where} \,\, 
\mathbf{H}_{j}(\mathbf{x}_j) = \zeta \left( \mathbf{W}_j \mathbf{x}_j + \mathbf{b}_j \right)
\label{eq:CPNN}
\end{equation}
In this equation, $\mathbf{y}$ is the CPNN output vector, $\mathbf{W}$ and $\mathbf{b}$ are the weight and bias of the $j^{th}$ layer, $m$ is the total number of layers, $\boldsymbol{\Theta}$ denotes the network's trainable parameters (i.e., weights and biases), $\zeta$ is the activation function, $\mathbf{H}$ is the $j^{th}$ (hidden) layer, $\mathbf{M}$ represents MIONet Eq. \ref{eq:MIONet}, and $\circ$ denotes a composition of functions, i.e., $\{\mathbf{H}_2 \circ \mathbf{H}_1 \}(\mathbf{x}) = \mathbf{H}_2(\mathbf{H}_1(\mathbf{x}))$.

\subsection{CPNN model training and evaluation} \label{training}

A database of 640 CFD simulations cases is split into training, validation, and test subsets for the CPNN model training and evaluation. Specifically, 512 cases (80\%) are used for training subsets, 64 cases (10\%) for validation subsets, and 64 cases (10\%) for test subsets. The validation subset is used for training regularization (e.g., early stopping checkpointing method) and hyperparameter optimization \citep{Prechelt2012}. The test subset is used to assess the model's predictive capability for unseen data. Before feeding through the CPNN, the input features of loading parameters $\mathbf{p}$, PM classes $\mathbf{cl}$, time stamps $\mathbf{T}$, and spatial coordinates $\mathbf{q}$ are first standardized. This involves centering the data by removing the mean and scaling to achieve unit variance. As illustrated in Fig. \ref{fig:architecture}, these standardized inputs $\mathbf{p}$, $\mathbf{cl}$, $\mathbf{T}$, and $\mathbf{q}$ are transformed into multidimensional tensors. For example, during training, the input tensors have dimensions of $(N_{L}, 1, 1, 1, 5)$ for loading parameters $\mathbf{p}$, $(1, N_{c}, 1, 1, 1)$ for PM class $\mathbf{cl}$, $(1, 1, N_{t}, 1, 1)$ for time stamps $\mathbf{T}$ and $(N_{L}, 1, 1, N_{s}, 3)$ for spatial coordinates $\mathbf{q}$, where $N_{L}$ is the number of training cases, $N_{c}$ is the number of particle classes, $N_{t}$ is the number of time samples, and $N_{s}$ is the number of spatial samples. These dimensions differ because each input type occupies a distinct axis to facilitate broadcasting and merging in MIONet. In this configuration, the loading parameters, the PM class, and the time stamps occupy the first, second, and third feature axes, respectively. Moreover, only a subset (8,000 spatial nodes) of the total CFD meshing nodes is used as spatial samples to train the NN. To ensure that the model is trained on the entire geometrical domain, this spatial sampling varies between cases rather than being duplicated. As a result, the spatial nodes occupy two feature axes: the first axis for the total number of loading cases , and the fourth axis for the number of spatial samples. Furthermore, the fifth (final) feature axis represents the number of features in each input group.

In this study, a mini-batch approach is employed to (1) minimize the graphics processing unit (GPU) memory usage (this will be discussed in detail later on), (2) accelerate the ML training process by calculating gradients from a portion of the training data rather than the full dataset, (3) implement a form of regularization in stochastic gradient descent, which enhances the model's ability to generalize \citep{Masters2018}. The framework is implemented in PyTorch \citep{Li2024a}. Mixed-precision training (Float16 and Float32) is adopted to further accelerate computation and lower GPU memory usage. At the start of each training iteration, a total of $N_L^{b}$ loading cases, $N_c^{b}$ PM classes, $N_t^{b}$ time stamps, and $N_s^{b}$ spatial points are randomly selected from the dataset to form a mini-batch, which is fed into the CPNN. In the CPNN forward pass, the branch networks compute the parameters $\mathbf{p}$ and $\mathbf{cl}$ across $N_l^e$ hidden layers containing $N_h$ neurons each, generating latent vectors $\mathbf{Br1}$ and $\mathbf{Br2}$ with dimensions $(N_L^{b}, 1, 1, 1, N_h)$ and $(1, N_c^{b}, 1, 1, N_h)$. Concurrently, the trunk networks manage the data for $\mathbf{T}$ and $\mathbf{q}$, converting them into latent vectors $\mathbf{Tr1}$ and $\mathbf{Tr2}$ with dimensions $(1, 1, N_t^{b}, 1, N_h)$ and $(N_L^{b}, 1, 1, N_s^{b}, N_h)$. These vectors are then merged using the Hadamard product $\mathbf{Tr1} \odot \mathbf{Tr2} \odot \mathbf{Br1} \odot \mathbf{Br2}$ to produce a time-position-encoded latent vector of dimensions $(N_L^{b}, N_c^{b}, N_t^{b}, N_s^{b}, N_h)$. This latent vector is subsequently passed through $N_l^f$ layers of the FCNN, resulting in the final output vector $(N_L^{b}, N_c^{b}, N_t^{b}, N_s^{b}, N_o)$, where $N_o$ is the number of outputs. In this study, there are two main outputs: PM concentration $c$ and velocity magnitude $|\mathbf{u}|$. A separate NN is used to predict each output, and hence $ N_o = 1$ for each of them. An alternative approach is to use the same NN to predict both outputs simultaneously. Both methods are implemented and tested. The results show that the two approaches yield comparable results for predicting velocity magnitude $|\mathbf{u}|$, as shown in Fig. S7 in the online Supplemental Materials and Fig. \ref{fig:c_|u|_standarded_evaluation}. However, the approach of using a separate NN for predicting $c$ yields better performance. Furthermore, this approach requires a smaller memory footprint. Therefore, the approach of separating NN for each output is adopted. This training strategy also aligns with the current operator learning research \citep{Cai2021}, where multiple NNs are implemented, with each NN predicting one output. Moreover, various NNs are trained on different particle classes to evaluate their sensitivity to loading parameters, as shown in Section \ref{Evaluation}. For position encoding, the Hadamard product $\mathbf{T} \odot \mathbf{B}$ is not the sole option; an element-wise addition $\mathbf{T} + \mathbf{B}$ can also be employed \citep{Tancik2020, Jin2022, Li2024a}. Finally, if a full-batch approach were adopted, then $N_{L}$, $N_{t}$, and $N_{s}$ would correspond to $M$ storm events, $O$ time instances, and $N$ spatial samples instead of $N_{L}^{b}$, $N_{t}^{b}$, and $N_{s}^{b}$, respectively.  

\section{Results}

\subsection{Comparison of ANN, MIONet and CPNN for predicting HS hydrodynamics and PM transport} \label{sec:model_comparison}

\begin{figure}[H]
\centering
\includegraphics[width=1.0\textwidth]{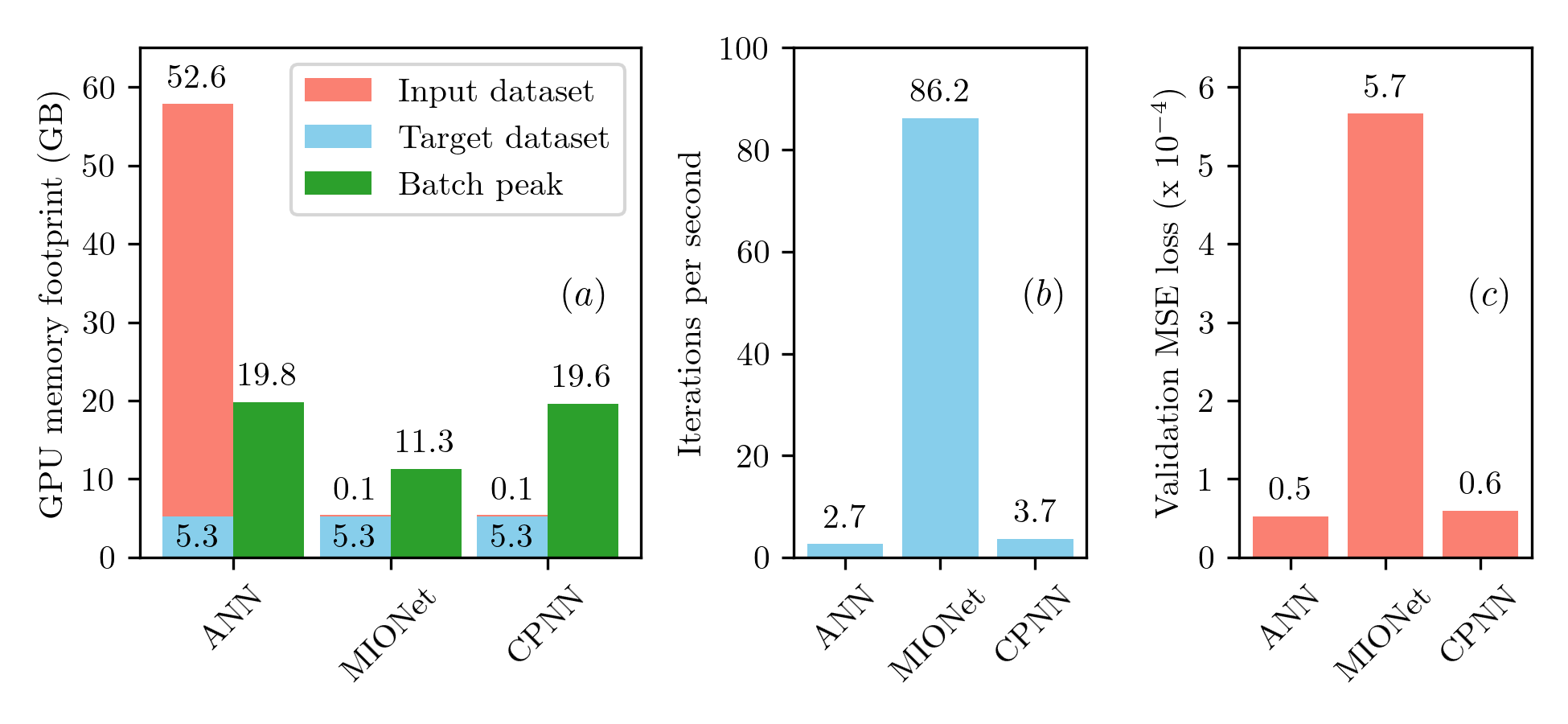}
\caption{Comparison of baseline models ANN, MIONet, and CPNN for predicting PM concentration. (a) GPU memory footprint for the full input dataset, the target dataset, and the peak usage of a single mini-batch during training, (b) Training speed measured by iterations per second. Each model consists of approximately 365,000 trainable parameters, and (c) Validation errors measured by the standardized mean squared error $\mathrm{MSE}$. These errors are computed for PM concentration $c$. Detailed model configurations are provided in Table S2 in the online Supplemental Materials.}
\label{fig:baseline_comparison}
\end{figure}

Fig. \ref{fig:baseline_comparison} compares the predictive capabilities of the baseline models ANN, MIONet, and CPNN for predicting PM concentration $c$. These baseline models have approximately the same number of trainable parameters (365,000), and their hyperparameters have not yet been optimized. The detailed configurations (e.g., number of layers and hidden neurons) are provided in Table S2 in the online Supplemental Materials. These models are trained with identical batch sizes using the Adam optimizer for 40,000 iterations, after which validation losses for velocity magnitude $|\mathbf{u}|$ and PM concentration $c$ plateau, indicating convergence, as shown in Fig. S5 in the online Supplemental Materials.

Fig. \ref{fig:baseline_comparison}a illustrates the memory footprints of the full input and target datasets in the three models. The ANN shows significantly higher VRAM usage than the total input dataset generated from the separate input groups. This is because ANN repeats spatial coordinates ($\mathbf{q}$), time stamps ($\mathbf{T}$), and PM class ($\mathbf{cl}$) for each loading case ($\mathbf{p}$), forming a much larger input size, as discussed in Section \ref{operator}. In contrast, MIONet and CPNN process input groups separately, eliminating this redundancy. As a result, the total input size is significantly reduced from 52.6 GB to 0.05 GB (approximately 0.08\% of the ANN input in this case), as shown in Section \ref{training}. This reduction in VRAM usage is critical for ML applications handling large datasets, such as the 3D time-dependent HS system presented here. Additionally, Fig. \ref{fig:baseline_comparison}a shows that the total target size remains unchanged across all models, as the output data has no redundancy. The figure also presents peak GPU memory usage (maximum VRAM allocation) for a single batch using the mini-batch method during training. This value includes the small memory footprint of the batch's input and target tensors, but excludes the memory used to store the total input and output datasets (preloaded into VRAM to accelerate batch sampling). The peak GPU memory usage reflects the maximum cumulative storage of intermediate activations, gradient buffers, and other temporary data required for the forward pass and backpropagation. This footprint mainly depends on the mini-batch sizes, the number of NN layers, and the number of hidden neurons per layer. In all models, a single mini-batch requires a peak memory footprint that exceeds 10 GB. The primary memory cost comes from processing the expanded tensor of shape $(N_L^{b}, N_c^{b}, N_t^{b}, N_s^{b}, N_h)$. ANN and CPNN exhibit comparable peak memory usage because both models process this tensor through the FCNN. In ANN, this large tensor is generated in the first layer and retains the same dimensions until the tensor reaches the final layer, where it is then transformed into the target dimensions. In contrast, the CPNN forms this expanded tensor after applying the Hadamard product to the outputs from its trunk and branch networks, then processes it through subsequent FCNN layers. Thus, although the timing of the tensor formation differs between the two models, the overall mini-batch memory footprint peak is similar and highly dependent on the number of layers and hidden neurons in the FCNN that process this expanded tensor. On the other hand, MIONet demonstrates lower memory consumption than ANN and CPNN. This is because MIONet mainly consists of trunk and branch networks, whose inputs and intermediate outputs are smaller. The peak memory usage in MIONet occurs in the final stage of the forward pass, when PyTorch broadcasts the outputs from these networks and combines them via the Hadamard product to produce the NN output.

Fig. \ref{fig:baseline_comparison}b compares the training speed of each model, showing that MIONet is significantly faster than ANN and CPNN. This is because MIONet processes the smaller tensors that have unity in most of their feature axes (($N_{L}, 1, 1, 1, ..$), ($1, N_{c}, 1, 1, ..$), ($1, 1, N_{t}, 1, ..$), ($N_{L}, 1, 1, N_{s}, ..$)), While ANN and the FCNN part of CPNN take considerable time to process the substantially larger tensor ($N_L^{b}, N_c^{b}, N_t^{b}, N_s^{b}, ..$). MIONet training speed is significantly higher than that of the DeepONet reported in our previous study \citep{Li2024a}. This could be attributed to MIONet's higher efficiency in removing redundancy, as discussed in Section \ref{operator}, and the use of only half as many layers compared to the earlier DeepONet study. By contrast, the CPNN trains more slowly than in our previous work because the total mini-batch size (batch case number, batch time sampling size, batch spatial points sampling size) is approximately 100 times larger. This decreases the speed of CPNN, particularly in the FCNN. Ultimately, Fig. \ref{fig:baseline_comparison}c shows that the MIONet architecture yields lower performance compared to ANN and CPNN when tested on validation data. For scenarios with a higher degree of nonlinearity like the ones presented here, the NN may need the FCNN to capture the nonlinear interactions and couplings between the system input groups (i.e., event loading parameters $\mathbf{p}$, PM class $\mathbf{cl}$, time stamps $\mathbf{T}$, and continuous coordinates $\mathbf{q}$). The MIONet architecture works well in situations where the solution does not require the expressivity of the FCNN to capture interactions between the inputs \citep{Jin2022}. In the MIONet, the four input parameter groups are independently processed through separate NNs, with their interaction taken into account only in the final Hadamard product operation. MIONet can lack the expressive capacity to account for the nonlinear interaction introduced by the FCNN (decoding part) in the CPNN. Adding the FCNN significantly improved the expressivity of the model, achieving a standardized mean squared error $\mathrm{MSE}$ of approximately $0.6\times10^{-4}$ on validation cases compared to $5.7\times10^{-4}$ in MIONet, as shown in Fig. \ref{fig:baseline_comparison}c.

\subsection{Influence of model architecture and training hyperparameter tuning}

\begin{figure}[H]
\centering
\includegraphics[width=1.0\textwidth]{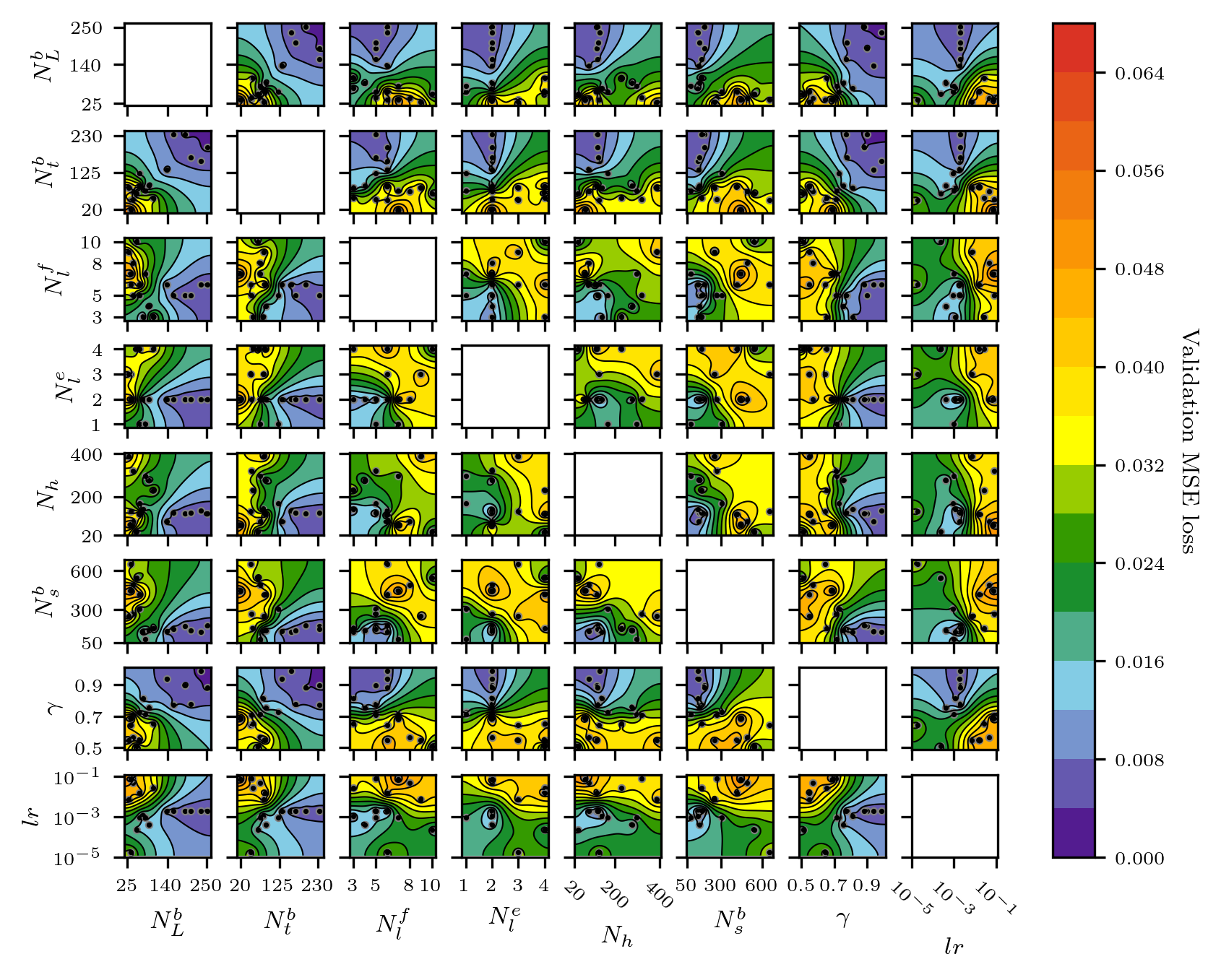}
\caption{Response surface of CPNN performance in Bayesian optimization. The model's predictive capability is measured by the validation $\mathrm{MSE}$ loss $\mathcal{L}_v$. $N_L^{b}$, $N_t^{b}$, $N_s^{b}$, $N_l^e$, $N_l^f$, $\gamma$, $lr$ are batch case number, batch time sampling size, batch spatial points sampling size, number of hidden neurons, number of encoding layers, number of FCNN layers, decay rate, and model learning rate. A total of 140 trials are performed, each with 25,000 training iterations. $\mathrm{MSE}$ loss is calculated based on standardized values. Only a subset of the full range of trial parameters is plotted for better visualization. The trial results are provided in the online Supplemental Materials Tables S3, S4, and S5.}
\label{fig:extended_optuna_standardized}
\end{figure}

Fig. \ref{fig:extended_optuna_standardized} shows the response surface of the validation $\mathrm{MSE}$ loss $\mathcal{L}_v$ with respect to the configuration parameters, including the batch case number ($N_L^{b}$), PM class sampling size ($N_c^{b}$), batch time sampling size ($N_t^{b}$), batch spatial points sampling size ($N_s^{b}$), number of hidden neurons ($N_h$), number of MIONet encoding layers ($N_l^e$), number of FCNN layers ($N_l^f$), learning rate decay ($\gamma$), and learning rate ($lr$). 140 trials are conducted using a Bayesian-based optimization algorithm, Optuna's Tree-structured Parzen Estimator (TPE) \citep{Akiba2019}. Fig. \ref{fig:extended_optuna_standardized} shows a subset of the full range of trials parameters. Complete results for all Optuna trials are provided in the online Supplemental Materials Tables S3-S5. In each trial, the model is trained for 25,000 iterations. 

The architecture depth, specifically the number of encoding layers $N_l^e$ and decoding layers $N_l^f$ plays a crucial role in learning the system dynamics. The results indicate a preference for a lower number of encoding layers and a relatively deeper FCNN layers configuration. One or two encoding layers often suffice, with an optimal depth of 4 to 7 FCNN layers. This configuration is supported by previous findings \citep{Li2024a}, which shows a deeper FCNN in effectively capturing interactions between the input features, time, and spatial coordinates. Increasing the number of encoding layers tends to negatively affect performance, as evidenced by the contour plot of $N_l^e$. This result is consistent with the findings reported in our earlier study. Furthermore, the optimal performance typically occurs with a narrow learning rate $lr$ range around 0.002 and a high decay rate $\gamma$ of 0.9.

The model performance exhibits a noticeable dependence on the batch case number $N_L^{b}$. Both extremes in batch case number, such as full-batch training and trivial mini-batches, yield inferior model performance. The ideal range for $N_L^{b}$ is found to lie between 140 and 250. This response pattern is mirrored in the batch time sampling size $N_t^{b}$. The ideal range for $N_t^{b}$ falls within 140 and 230. For batch spatial points sampling size $N_s^{b}$, it is remarkable to find that the optimal range is very low (less than 3\% of the full-batch). The optimal range is found to be between 80 and 250, which is significantly lower than the 8,000 coordinate sampling points in the dataset. 

The ideal range for $N_h$ is found to be between 90 and 120 neurons. This is considerably lower than the 500 neurons per layer reported in \citep{Li2024a}. The number of neurons controls the representation of the input feature space. Consequently, the lower optimal number of neurons observed here may suggest that transient loadings for this HS system require fewer neurons for effective representation than the basin's geometrical variations studied previously. This is advantageous since the number of neurons $N_h$ dictates the peak size of the batch GPU memory footprint, as explained in Section \ref{sec:model_comparison}. Additionally, using fewer hidden neurons per layer helps prevent the model from overfitting the training data. Ultimately, the optimized model's hyperparameters are $N_L^{b}$ = 226, $N_t^{b}$ = 227, $N_s^{b}$ = 195, $N_l^e$ = 2, $N_l^f$ = 6, $N_{h}$ = 92, $\gamma$ = 0.984, and $lr$ = 0.002. This optimized configuration is adopted in the subsequent evaluation.

\subsection{Evaluating CPNN predictive capability and generalizability}
\label{Evaluation}

\begin{figure}[H]
\centering
\includegraphics[width=1.0\textwidth]{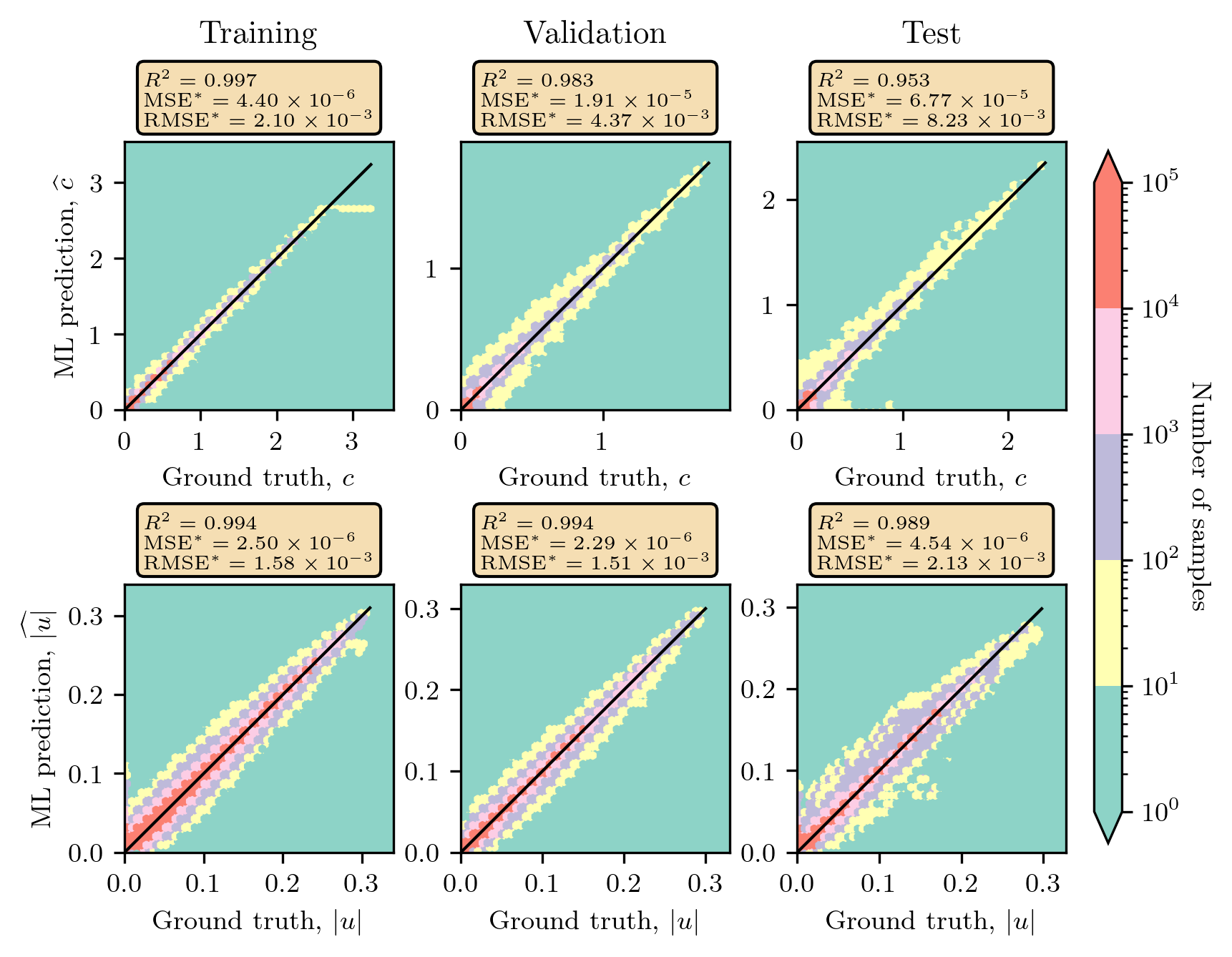}
\caption{Model performance on training, validation, and test datasets for both the concentration $c$ and the velocity magnitude $|\mathbf{u}|$. The dataset consists of 640 cases:  512 cases (80\%) for training, 64 cases (10\%) for validation, and 64 cases (10\%) for test. The colormap shows the number of predicted points at each location on the plot. The black diagonal line indicates the perfect prediction. $R^2$, $\mathrm{MSE}^\ast$, and $\mathrm{RMSE}^\ast$ are the coefficient of determination, mean-square error, and root-mean-square error and are calculated based on the unstandardized field values.}
\label{fig:c_|u|_standarded_evaluation}
\end{figure}

Fig. \ref{fig:c_|u|_standarded_evaluation} illustrates the model's predictive capability across training, validation, and test datasets for PM concentration $c$ and velocity magnitude $|\mathbf{u}|$. Overall, CPNN demonstrates higher predictive performance for modeling hydrodynamics and PM transport. The coefficient of determination $R^2$ values range from 0.953 to 0.997 for concentration and from 0.989 to 0.994 for velocity magnitude, indicating an excellent fit between the predicted and actual values. The small reduction in predictive performance between the training and test datasets highlights the CPNN model's exceptional ability to generalize effectively.

\begin{figure}[H]
    \centering
    \includegraphics[width=1.0\textwidth]{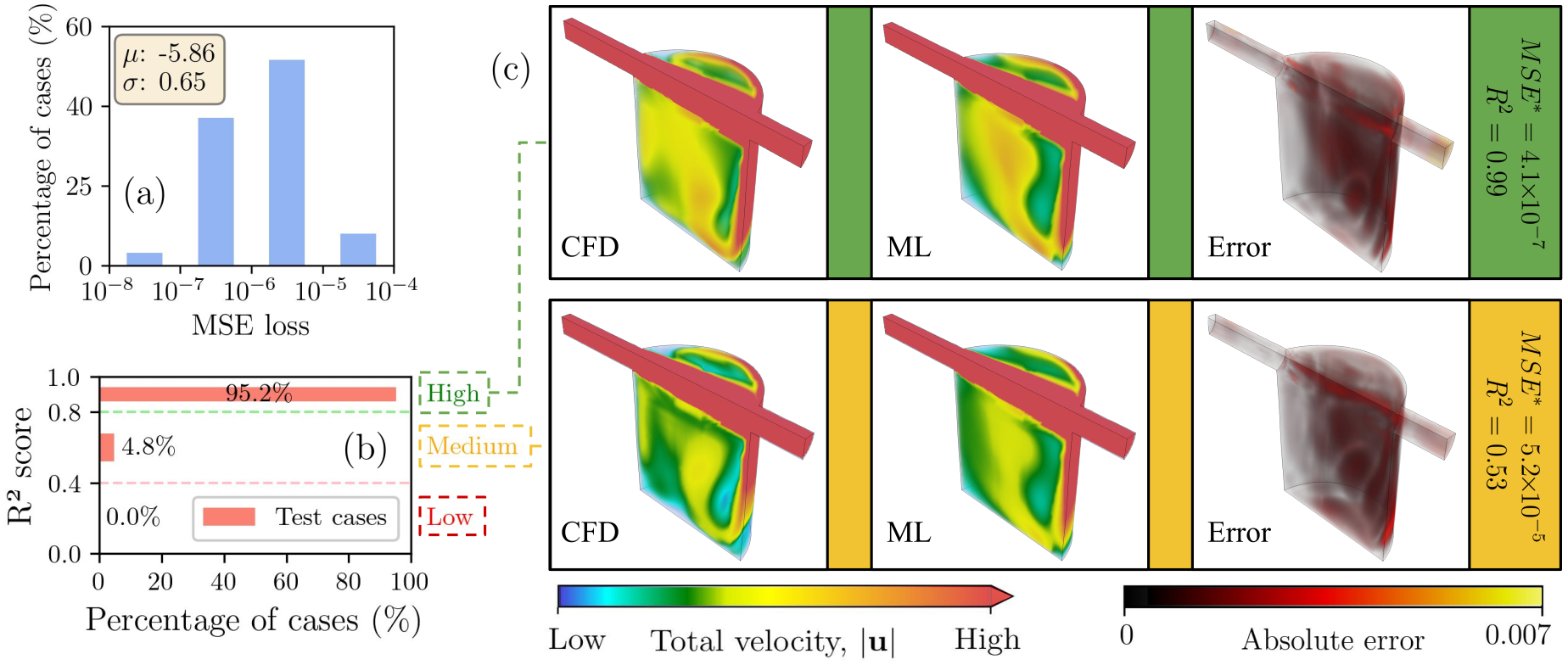}
    \caption{Model error distribution across different test cases for the velocity magnitude $|\mathbf{u}|$ prediction. (a) Unstandardized mean squared error $\mathrm{MSE}^\ast$ distribution, (b) $R^2$ distribution, and (c) Examples of different categories of model performance. The visualized 3D HS system is symmetrically clipped to highlight the dynamics in the system core. Velocity magnitude is capped below the maximum to better visualize fine circulation flow. The test dataset includes 64 cases. $\mu$ and $\sigma$ are the mean and standard deviation of the log-normal distribution with a base of 10. $\mathrm{MSE}^\ast$ and $R^2$ are calculated based on the unstandardized field values.}
    \label{fig:V_error}
\end{figure}

While Fig. \ref{fig:c_|u|_standarded_evaluation} provides the overall evaluation of the CPNN generalizability, these results do not explore the model's performance on an individual test case basis. To gain a deeper understanding of the model's generalizability, the performance of the CPNN model is analyzed separately for each test case. Each case's unstandardized mean-square error $\mathrm{MSE}^\ast$ and $R^2$ values are computed by averaging the error across all the case's solution points. When evaluating the velocity magnitude $|\mathbf{u}|$ prediction, Fig. \ref{fig:V_error}a suggests that the velocity magnitude $\mathrm{MSE}^\ast$ loss generally follows a log-normal distribution with $\mu$ = -5.86 and $\sigma$ = 0.65 (logarithm base 10). Fig. \ref{fig:V_error}b shows that 95.2\% of the test cases have been classified in the high prediction accuracy category ($R^2>0.8$). Only 4.8\% of the cases fall under the medium category, while none of the cases is classified as a low category case. These results demonstrate the feasibility of higher-fidelity hydrodynamics predictions with CPNN. Nevertheless, despite these high $R^2$ scores, they do not guarantee that CPNN captures all the fine details of the flow. Visual comparison of the model and CFD results indicates that CPNN can predict dominant jet flow from the inlet to the outlet, while some fine details of the circulating flow are missing, as shown in Fig. \ref{fig:V_error}c (the velocity magnitude is capped below the maximum in this figure to better visualize fine circulation). These findings will be discussed in detail in the Discussion Section. 

\begin{figure}[H]
\centering
\includegraphics[width=1.0\textwidth]{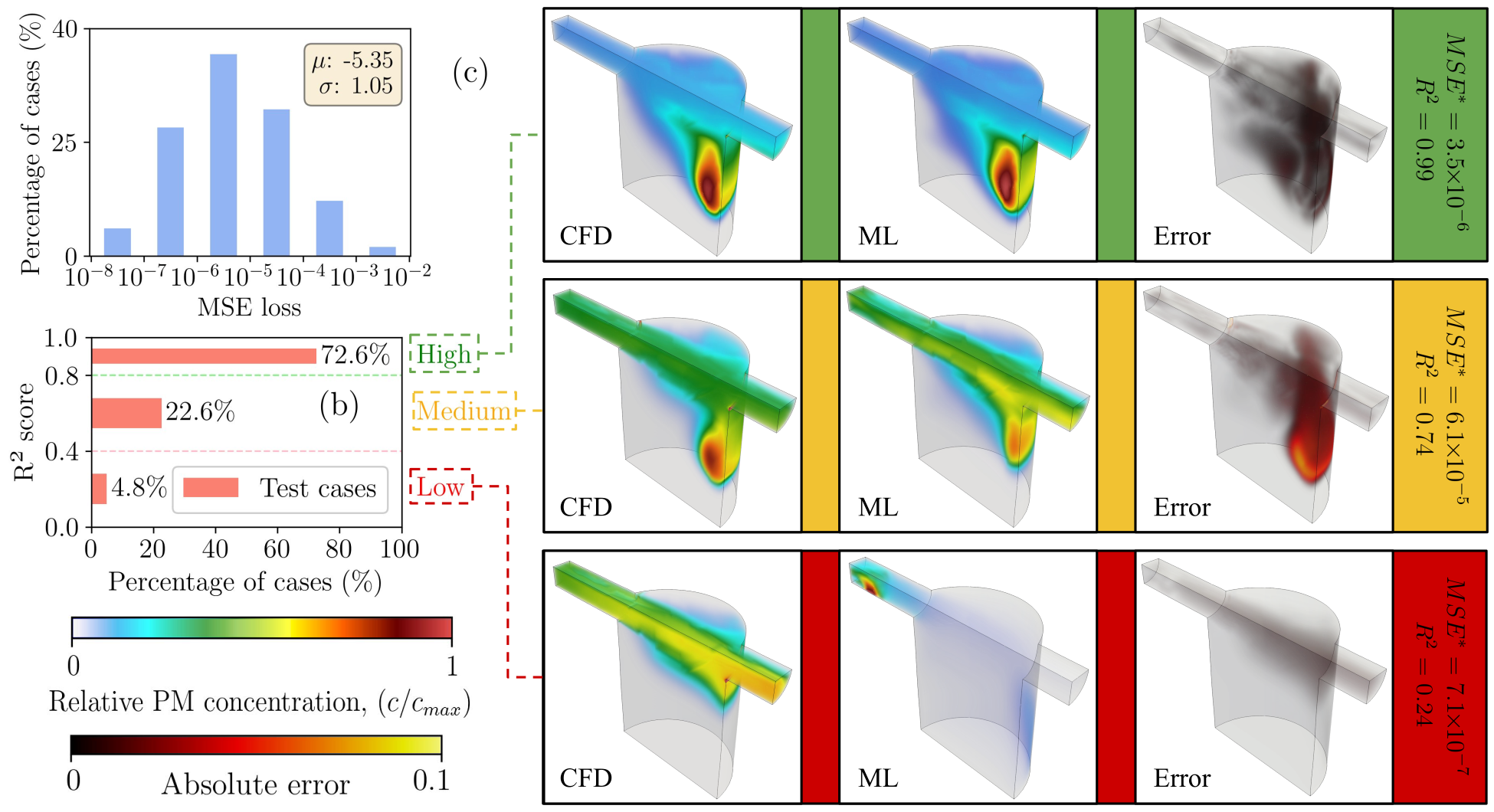}
\caption{Model error distribution across different test cases for the concentration $c$ prediction. (a) $\mathrm{MSE}^\ast$ distribution, (b) $R^2$ distribution, and (c) Examples of different categories of model performance. Relative PM concentration is used to allow for consistent visualization across cases with varying absolute concentration ranges. The total number of test datasets is 64 cases. $\mu$ and $\sigma$ are the mean and standard deviation of the log-normal distribution with a base of 10. $\mathrm{MSE}^\ast$ and $R^2$ are calculated based on the unstandardized field values.}
\label{fig:C_error}
\end{figure}

Fig. \ref{fig:C_error} illustrates CPNN's performance for predicting the PM concentration $c$. A log-normal distribution of $\mathrm{MSE}^\ast$ with $\mu$ = -5.35 and $\sigma$ = 1.05 is observed. The CPNN model achieves higher prediction accuracy with an $R^2$ greater than 0.8 in approximately 72.6\% of cases, as shown in Fig. \ref{fig:C_error}b. Around 22.6\% of the test cases exhibit medium predictive capability $ 0.4 < R^2 < 0.8 $. Similar to the observed performance in velocity magnitude, CPNN captures the general pattern of PM transport, but misses some finer details. These discrepancies are primarily observed in the region where the inlet jet impinges on the outlet wall. The dispersion patterns of PM can be sensitive to the unsteady inlet loading conditions (hydrograph $Q_{\text{inlet}}$ and pollutograph $SSC_{\text{inlet}}$). This is likely due to the underlying nonlinear dynamics of turbulence transport inside the tank. Consequently, CPNN yields lower predicative capability in the regions of the PM dispersion. Among 4.8\% of the test cases, the CPNN model exhibits a poor predictive capability, with $R^2$ values below 0.4. Fig. \ref{fig:C_error}c shows that CPNN incorrectly predicts that most particles settle in the inlet pipe, whereas the CFD solution indicates particle dispersion and circulation in the tank. Such a discrepancy suggests that the CPNN underpredicts the inlet flow rate $Q_{inlet}$ in these cases, causing the inlet horizontal velocity component to be much lower than the particle's terminal velocity $w_s$, leading to premature PM settling. Further investigation of these cases revealed that the loading feature $\theta$ in these cases is subject to extremely low values ($<0.37$). Considering the $\theta$ range from 0.23 to 51.5 (Table \ref{tab:parameter_summary}), such an extreme $\theta$ value can challenge model learning. Furthermore, CPNN's Error convergence is investigated and the results are provided in Fig S8 in the online Supplemental Material.

\subsection{Sensitivity map of PM transport}
\label{Sensitivity}

One of the advantages of CPNN is its ability to compute the derivatives of the model's output (concentration $c$ and velocity magnitude $|\mathbf{u}|$) with respect to its input features. Not only can the NN compute spatial and temporal gradients (e.g., $\nabla \mathbf{u}$ or $\partial c/\partial t$), but it can also calculate solution derivatives with respect to the storm event loading parameters. This allows the CPNN to evaluate the impact of changing any loading parameter on the PM transport in the 3D system at any given time. 

\begin{figure}[H]
\centering
\includegraphics[width=1.0\textwidth]{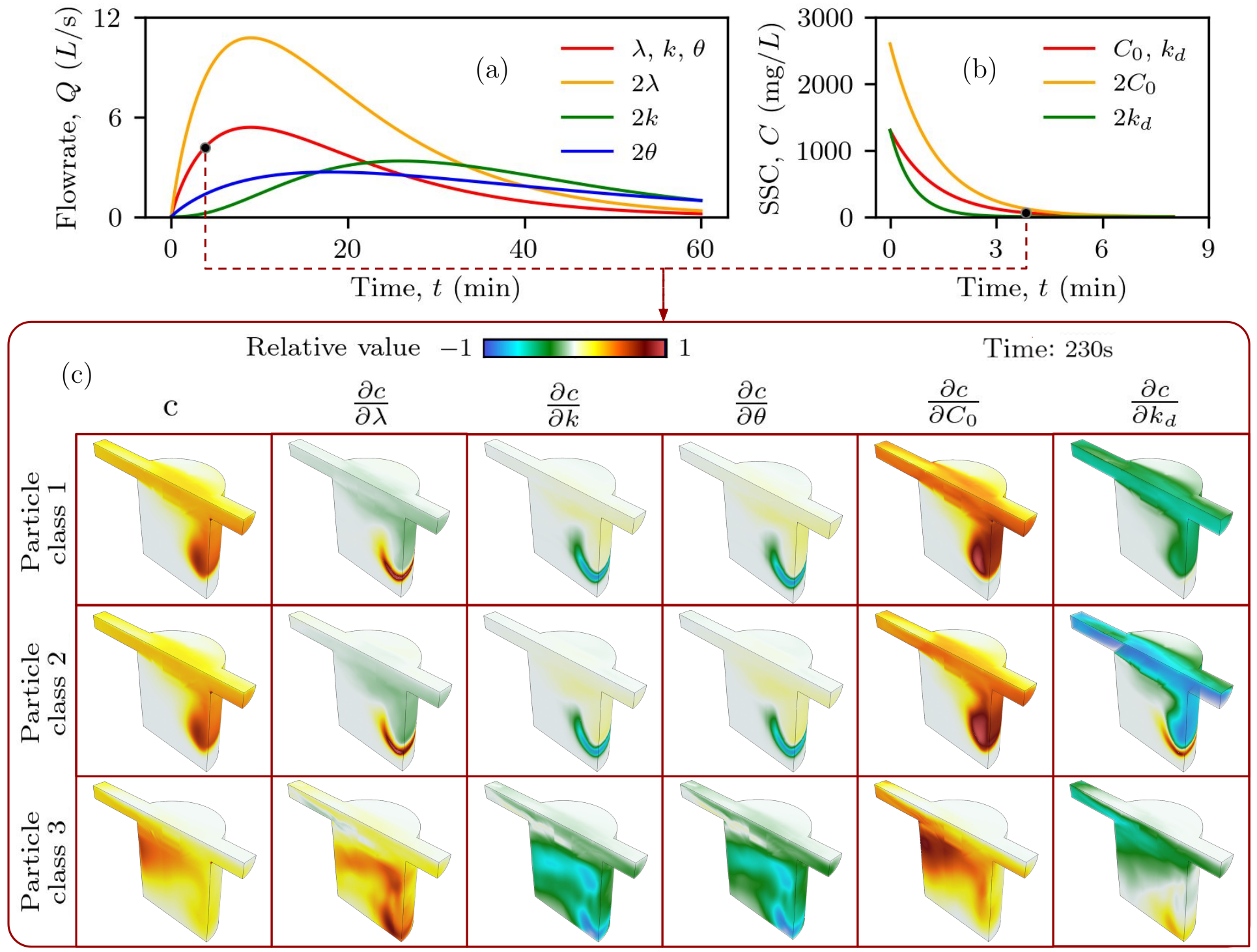}
\caption{Dependencies of the HS system's PM concentration $c$ on the storm event's loading parameters for a test case. (a) The selected stormwater event's hydrograph ($\lambda$ = 0.14, $k$ = 1.9, and $\theta$ = 10) as a baseline. Additionally, the figure shows hydrographs generated by doubling the value of each parameter($\lambda$, $k$, and $\theta$), (b) The stormwater event's pollutograph ($C_0$ = 1.3, and $k_d$ = 0.8) as a baseline. Moreover, the figure shows pollutographs generated by doubling the value of any of its parameters ($C_0$, and $k_d$, and (c) 3D symmetrically clipped contours of the system at a specific time (230s) during the stormwater event. The first column represents the concentration contour. The following columns represent the concentration gradient mapping with respect to the different loading conditions. The three rows represent different particle classes that go into the system. Each class is associated with a different terminal velocity (${w_s}^{d1} = 7.5 \times 10^{-5} m/s$, ${w_s}^{d2} = 3.16 \times 10^{-4} m/s$, and ${w_s}^{d3} = 5.62 \times 10^{-3} m/s$). Relative value is the value at each node divided by the maximum absolute value in the entire domain during the event for each contour.}
\label{fig:sensitivity_map}
\end{figure}

Since this problem is high-dimensional, a case with specific $\lambda$, $k$, $\theta$, $C_0$, and $k_d$ has to be chosen to calculate its gradients. Fig. \ref{fig:sensitivity_map}a and b show the hydrograph and pollutograph of the selected test case. This baseline case has the loading parameters $\lambda$ = 0.14, $k$ = 1.9, $\theta$ = 10, $C_0$ = 1.3, and $k_d$ = 0.8. Additionally, Fig. \ref{fig:sensitivity_map}a and b show the effect of doubling the value of any of these loading parameters on the resulting hydrograph and pollutograph. These graphs illustrate the influence of each parameter on the solution space and establish a general expectation of the gradient mapping contours shown in Fig. \ref{fig:sensitivity_map}c.

Furthermore, Fig. \ref{fig:sensitivity_map}c visualizes the 3D symmetrically clipped contours at a certain point in time (230$s$). The first column shows the actual concentration, while the following columns show the concentration derivative with respect to each loading parameter, such as $\partial c/\partial \lambda$. Each row shows the contours of a different particle class in order to provide deeper insights into each class's dynamics and their dependencies on the inlet loading conditions in the HS system. Three particle classes are compared in the figure, each associated with a terminal velocity (${w_s}^{d1} = 7.5 \times 10^{-5} m/s$, ${w_s}^{d2} = 3.16 \times 10^{-4} m/s$, and ${w_s}^{d3} = 5.62 \times 10^{-3} m/s$). Classes with higher terminal velocities settle more rapidly. Video S1 (online Supplemental Materials) shows the evolution of the gradient mapping during the chosen storm event. 

Fig. \ref{fig:sensitivity_map}c shows the effects of the hydrograph loading parameters ($\lambda$, $k$, and $\theta$) on the first flush. For example, the concentration derivative with respect to $\lambda$ has a positive relative value in the first flush leading edge, and its region of effect goes beyond the actual concentration area, as shown for the ${w_s}^{d3}$ class. This shows that if $\lambda$ had a higher value, the leading edge of the first flush would have settled closer to the bottom of the system by the selected time point. This is physical since higher $\lambda$ leads to higher flow rate $Q$ going into the system, as shown in Fig. \ref{fig:sensitivity_map}a, and results in an increase in the first flush's velocity, filling more area with PM. Additionally, the ${w_s}^{d3}$ class contour shows a higher positive derivative near the bottom and near the outlet walls, which further aligns with the expectation of a higher inlet flow rate $Q$, increasing PM horizontal transport. Some of the later portions of the transported PM exhibit a negative relative derivative value. This is because dispersing the same amount of PM in a larger volume decreases the concentration value in these areas compared to the previously slow and highly condensed case. Additionally, Fig. \ref{fig:sensitivity_map}a shows that increasing $k$ decreases the hydrograph peak, makes the hydrograph more symmetric, and shifts it to the right. This leaves an initial period with nearly zero inflow.  As a result of the lower and delayed peak, not only should the concentration values decrease, but the shift in hydrograph should also cause the first flush to arrive later. The visualization of the derivative with respect to $k$ in Fig. \ref{fig:sensitivity_map}c aligns well with the expected behavior. The leading portion of the first flush has a negative relative value, indicating that this area would have had zero concentration at the time if the loading parameter $k$ was higher. Some of the later portions of the transported PM have a positive relative derivative value. This is because dispersing the same amount of PM in a smaller volume increases the concentration value in these areas compared to the previously quick and stretched case, opposite to the $\lambda$ effect. In addition, Fig. \ref{fig:sensitivity_map}a shows that doubling the $\theta$ value decreases the hydrograph peak more than doubling the $k$ value, while also shifting the peak to the right. Fig. \ref{fig:sensitivity_map}c shows that the leading portion of the first flush concentration would decrease by increasing $\theta$, which is relatively similar to the $k$ effect on the concentration in the system. 

The pollutograph loading parameters ($C_0$ and $k_d$) also influence the first flush PM concentration. Increasing $C_0$ results in the graph starting from a higher concentration value, as shown in Fig. \ref{fig:sensitivity_map}b. As a result, the concentration rises in regions that previously had dispersed PM at the selected time. Unlike $\lambda$, which controls the flow rate, $C_0$ does not influence the flow velocity in the system. According to Fig. \ref{fig:sensitivity_map}b, increasing $k_d$ steepens the pollutograph's slope. This increases the magnitude of the negative slope while having no effect on the starting point. The derivative should have negative relative values in the same regions as the concentration contour to indicate a reduction in the PM concentration values. The first particle class agrees with the expectations. For the second and third particle classes, although the derivative is negative for most of the dispersed PM region, the first flush leading edge has a positive relative value. The ML model incorrectly predicts that an increase in $k_d$ affects the velocity of PM, prematurely dissipating pollutants in empty zones. This can be explained by comparing the influence of each of these derivatives. Although $C_0$ and $k_d$ have opposite effects on the system, doubling $C_0$ has a much higher effect on the system PM concentration than the effect of decreasing $k_d$ by half. This may justify why the model learned the derivatives $\lambda$, $k$, $\theta$, and $C_0$ more effectively than $k_d$. Due to its relatively small impact, the model demonstrates lower performance in learning the effect of $k_d$ compared to the other derivatives $\lambda$, $k$, $\theta$, and $C_0$. This is because supervised data-driven models don't learn the actual physics but correlations that approximate the result. Although the combined effect of these parameters results in higher predictive capability, there is no guarantee that the data-driven model has accurately learned the physical relationships between the inputs and outputs. Finally, this analysis is conducted on a test case and can be applied to other unseen scenarios to evaluate the system's response to new events.

\section{Discussion}

\subsection{Challenges in learning flow and PM solutions across orders of magnitude}
\label{orders of magnitude}

The CPNN model delivers robust overall predictions of 3D transient flow and PM fate dynamics, yet it underperforms on a small subset of cases. As shown in Fig. \ref{fig:V_error}, the CPNN yields $R^2 < 0.4$ for 4.8\% loading conditions. A close examination shows that in these less performing cases, the horizontal jet flow in the HS is orders of magnitude higher than the circulating flow, especially in the small inlet and outlet tubes. For example, in one of the test cases, the jet flow has a velocity magnitude of approximately 0.023 \unit{m/s}, while the circulating flow has a velocity magnitude of approximately 0.001 \unit{m/s}. This may explain why in the medium category cases, the NN is able to predict the jet flow with higher accuracy but had lower capability to predict the low-velocity-magnitude circulating flow. During training, the NN can prioritize reducing errors in regions with higher flow velocities, as mispredictions in these dominant areas contribute significantly to the overall $\mathrm{MSE}$. In contrast, errors in the low velocity magnitude circulating flow regions contribute minimally to the $\mathrm{MSE}$. Consequently, the NN predicts the fine circulating flow within a close range, resulting in a smaller absolute error but a higher relative error compared to the jet flow region predictions. 

A similar imbalance also arises in PM concentration $c$ predictions. Terminal velocity draws elevated PM downward, causing high concentrations to permeate the entire tank rather than remain confined to the inlet-outlet jet. As a result, both jet and core regions exhibit similar absolute‐error magnitudes. When the model underperforms on core concentrations, both $\mathrm{MSE}^\ast$ and $R^2$ worsen compared to the velocity predictions. Interestingly, Fig. \ref{fig:C_error}c shows that \quotes{low-category} cases sometimes can yield smaller absolute errors than medium or high categories. This counterintuitive result arises because these bad cases' PM concentration $c$ values span a much smaller range, so deviations remain small in absolute terms. As aforementioned, during training, the loss function prioritizes larger errors in medium‐ and high‐range samples, causing some low‐range patterns to be overlooked. In the case of concentration prediction, this leads to significantly lower predictive accuracy in capturing the main PM transport pattern, driven by the jet flow, when the case's range of absolute values is orders of magnitude lower than that of the other cases the NN is trained on. For the cases in the low category, although the NN mispredicts the concentration $c$ across most of the system, the $\mathrm{MSE}$ errors are overshadowed by those of the hundreds of medium- and high-range cases. Those higher-error cases took priority in updating the model's weights and biases. This highlights a challenge in learning from data that spans multiple orders of magnitude. A potential solution can be increasing the density of training data in the lower range of the solution space. This way, the model can be guided to learn critical behaviors across all magnitudes rather than predominantly focusing on higher absolute value cases. 

Additionally, this bias toward high‑magnitude cases can be mitigated by redefining the model's training domain. Currently, the network learns the solution of the inlet and outlet regions, which have much smaller cross‑sections, and thus substantially higher velocities, and an exceptionally high inlet concentration compared to the tank interior. As a result, the model spans a wider range of $|\mathbf{u}|$ and $c$, intensifying the difficulty of fitting solutions that span orders of magnitude. Excluding the inlet–outlet segments would eliminate low‑category concentration cases driven by particle settling in the inlet and remove extreme velocities, simplifying the learning of fine‑scale patterns. Nevertheless, we retain inlet and outlet regions to ensure the model's robustness and generalizability to other water infrastructure systems. Predicting outlet concentration also enables direct calculation of total suspended solids discharge, 
 which provides a critical metric of PM removal efficiency. Consequently, despite the added complexity, these regions are essential for both dynamic mapping and system performance evaluation.

\subsection{Enabling assessment of water infrastructure performance for a longer-term period}
\label{long-term}

\begin{figure}[H]
\centering
\includegraphics[width=1.0\textwidth]{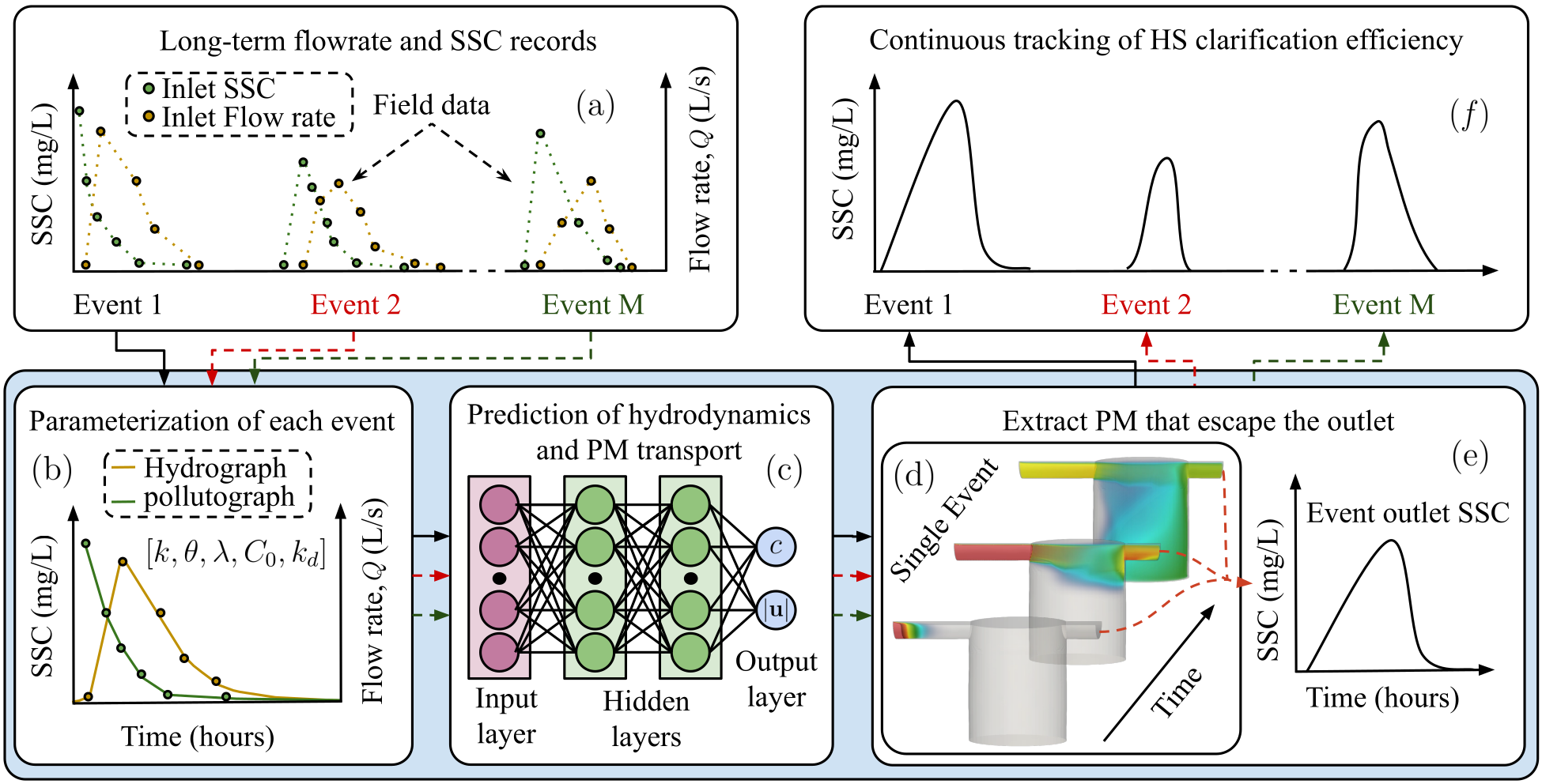}
\caption{Continuous evaluation of a stormwater system. (a) Long-term hydrograph and pollutograph records, (b) Parameterization of each event's hydrograph and pollutograph using modified gamma and exponential decay functions, (c) Forward pass of each event's parameters through the CPNN to get the output PM concentration $c$ and velocity magnitude $|\mathbf{u}|$, (d) 3D visualization of the system PM transport at different times during the event, (e) Accumulation of the predicted concentrations at the system outlet over the event period, which generates the event's SSC discharge graph, and (f) Consecutive aggregation of the individual events' SSC discharge graphs to form a continuous effluent concentration graph for the HS efficiency.}
\label{fig:Long_term_assessment}
\end{figure}

Assessing the long‑term performance of stormwater treatment systems is critical for resilient, cost‑effective urban planning and sustainable stormwater management. However, such evaluations remain challenging due to the high cost associated with long‑term field monitoring and the high computational cost of continuous CFD simulations. The CPNN framework developed in this study has the potential to overcome these barriers and enable efficient prediction of flow and PM transport over seasonal or multi‑year time horizons. Fig. \ref{fig:Long_term_assessment} illustrates a workflow with a CPNN-based framework. In this CPNN-based framework, the continuous hydraulic and PM loading records, whether from field monitoring or generated by a hydrological model such as SWMM, are first segmented into individual storm events , as illustrated in Fig. \ref{fig:Long_term_assessment}a. Each event is then parameterized using a modified gamma function for inflow and an exponential decay function for pollutant loading , as shown in Fig. \ref{fig:Long_term_assessment}b. These parameter sets are fed into the CPNN, which produces high‑resolution, transient 3D fields of velocity and PM concentration for each event (Fig. \ref{fig:Long_term_assessment}c). For each storm, the outlet concentration time series is integrated into the SSC discharge graph, as depicted in Figs. \ref{fig:Long_term_assessment}d and e. Finally, concatenating the per-event SSC results yields a continuous effluent concentration profile, providing a metric of system performance over extended periods, as shown in Fig. \ref{fig:Long_term_assessment}f. Furthermore, because the continuous hydraulic and PM‐loading record is decomposed into discrete storm events that the CPNN analyzes independently, the workflow can be parallelized and vectorized to exploit modern GPU hardware. 

Moreover, if two storms occur in rapid succession, without sufficient time for the system to drain, they should be treated as a single extended event, with intermediate zero‐flow and zero‐SSC intervals. While the modified gamma and exponential‐decay functions capture isolated hydrographs and pollutographs well, they struggle to represent multi‐peak or prolonged events. One remedy is to parameterize each event by a series of discretized time-concentration points, which maximizes flexibility but greatly increases input dimensionality and training‐data requirements. Future work should therefore explore a hybrid parameterization, combining a small set of analytic basis functions with targeted discretization, to balance expressive power against model complexity and training data demands.  

Beyond the HS system examined here, the CPNN framework offers broad applicability across diverse water infrastructure. By tailoring the CPNN inputs and retraining on system‑specific boundary conditions, the model can be extended to other water infrastructure, such as stormwater conveyance channels, retention basins, biological reactors, and even wastewater networks. ML-enabled higher‑fidelity transient 3D simulation remains nascent in this field, but a generalizable architecture like CPNN enables rapid, detailed prediction of nonlinear, unsteady hydraulics and pollutant transport across multiple system geometries, as shown in this study and our previous work \citep{Li2024a}. This versatility empowers engineers and decision‑makers to evaluate and compare performance under varied loading conditions, optimize designs, and plan operations and maintenance with greater confidence. Ultimately, CPNN helps pave the way toward more efficient, resilient water infrastructure capable of meeting both present needs and future challenges.

\section{Conclusion}

This study develops an operator-based machine learning framework, CPNN, for predicting the unsteady treatment dynamics of stormwater infrastructure. The predictive capability of this framework is benchmarked and demonstrated by applying it to a common urban drainage stormwater treatment system, the HS. The baseline models of CPNN, ANN, and MIONet are compared in terms of predictive capabilities, training efficiency, and GPU memory footprint. Furthermore, the influence of model parameters on CPNN performance is assessed. The performance of CPNN in predicting HS hydrodynamics and PM transport fate is evaluated against higher-fidelity CFD simulations. The findings of this study lead to the following conclusions:

\begin{itemize}

\item Modified gamma and exponential functions can effectively parameterize the stormwater infrastructure hydrograph and pollutograph. Regression analysis shows good agreement between the observed hydrograph and pollutograph over 40 events, as shown in Figs. S1-S4 in the online Supplemental Materials. Rather than representing the unsteady loading condition as a series of discrete point features, modified gamma and exponential functions characterize the loading condition using only five parameters. This significantly reduces the parameter space. Furthermore, in the CPNN model formulation, while the explicit mathematical expressions of the modified gamma and exponential functions are not directly incorporated, the model can effectively learn the inherent relationships governing system hydrodynamics and pollutant transport. Through training on parameterized input representations, CPNN can accurately capture the dependencies between these functions and key system responses. This allows for improved predictive performance across diverse unseen stormwater management scenarios. This capability underscores the advantage of operator-based machine learning frameworks in learning the underlying physics in the dataset.

\item The proposed CPNN framework leverages the strengths of both ANN and MIONet architectures. By segmenting the input dataset into the smallest separate input groups, CPNN eliminates redundancy and reduces the input dataset size significantly (by over 99\% for the current system) compared to standard ANNs. Furthermore, the integration of MIONet with an FCNN enhances the model's capacity to resolve the nonlinear interactions between loading parameters, PM classes, time stamps, and spatial coordinates. The CPNN improves upon MIONet ($\mathrm{MSE}^\ast$ of $5.7\times10^{-4}$) and achieves an $\mathrm{MSE}^\ast$ of $6\times10^{-5}$ on the validation dataset as shown in Fig. \ref{fig:baseline_comparison}.

\item The CPNN model demonstrates strong generalizability in predicting HS flow hydrodynamics and PM transport under unsteady loading conditions. Specifically, the $R^2$ for PM concentration in test cases ranges from 0.953 to 0.997, while for velocity magnitude $|\mathbf{u}|$, it ranges from 0.989 to 0.994. A case-based evaluation shows that for velocity magnitude predictions, the model achieves higher accuracy in 95.2\% of test cases and medium accuracy in the remaining 4.8\%, with no cases falling into the lower accuracy category. For PM concentration predictions, approximately 72.6\% of cases are classified as higher accuracy, 22.6\% fall within the medium accuracy range, and only 4.8\% exhibit lower accuracy, as shown in Figs. \ref{fig:V_error} and \ref{fig:C_error}.

\item The unique automatic differentiation capability of CPNN provides an efficient approach for gradient computation and evaluating system sensitivity. The sensitivity analysis with respect to loading parameters reveals that scaling factor $\lambda$ strongly accelerates the first flush, pushing PM deeper into the system earlier, while higher shape parameter $k$ and scale parameter $\theta$ delay or dampen that flush, reducing early‐stage PM build‐up. Doubling initial concentration parameter $C_0$ elevates PM concentrations wherever flow already exists, suggesting a higher system‐wide sensitivity to initial pollutant levels. By comparison, changes in decay concentration coefficient $k_d$ have a smaller overall impact, as shown in Fig. \ref{fig:sensitivity_map}.

\item Challenges remain for CPNN in learning finer hydrodynamic details and handling extreme loading conditions at the lower end of the solution space, particularly when the loading conditions span multiple orders of magnitude. Due to the $\mathrm{MSE}$-based loss function, storm events with lower PM concentrations produce smaller absolute errors, which may be outweighed by cases with higher solution ranges. As a result, the model has a reduced \quotes{incentive} to accurately learn low-magnitude scenarios. These findings highlight the need to increase the sampling density of training data in the lower ranges of the solution space, allowing the model to better capture low-end patterns without predominantly focusing on high-end scenarios, thereby improving accuracy for extreme loading conditions.

\item The framework has potential for the long-term performance evaluation of water infrastructure systems. By segmenting long-term hydrological records into individual storm events and feeding them into the CPNN, the model can predict the system's unsteady response for each event. The transient effluent concentration predictions can then be aggregated to generate the corresponding SSC discharge graph throughout each of these events. This can enable a continuous assessment of the system's treatment performance over a significantly longer period of time. 

\end{itemize}

\section*{Acknowledgments}
This research was supported by the AI Tennessee Initiative and by UT-Oak Ridge Innovation Institute at the University of Tennessee.

\section*{Supplemental materials}
Figs S1-8, Tables S1-6, and Video S1 are available online.

\section*{List of symbols}
\setlength\tabcolsep{3pt}
\renewcommand{\arraystretch}{1.0}
\begin{tabular}{llll}
\toprule
$Q(t)$ & Inlet flow rate (time-dependent) 
    & $w_s$ & Particle terminal velocity\\
$d$ & Particle diameter 
    & $\lambda$ & modified gamma function scaling factor \\
$k$ & modified gamma function shape parameter 
    & $\theta$ & modified gamma function scale parameter \\
$C_0$ & Initial pollutograph concentration 
    & $k_d$ & Pollutograph decay coefficient \\

$\mathbf{p}=[\lambda,k,\theta,C_0,k_d]$ 
    & Storm-event parameters 
    & $\mathbf{cl} = [w_s]$ & Particle class (terminal velocity)  \\
$\mathbf{q} = [x,y,z]$ 
    & Spatial coordinates 
    & $\mathbf{T} = [t]$ & Time stamps \\
$M$ & Number of total cases 
    & $M_t$ & Number of training cases \\
$O$ & Number of total time snapshots 
    & $N$ & Number of total spatial sample points \\
$\mathbf{x}$ & Generic input 
    & $\mathbf{y}$ & Generic output \\
$\mathbb{X}$ & Global input matrix 
    & $\mathbb{Y}$ & Global solution matrix \\
$c$ & PM concentration 
    & $|\mathbf{u}|$ & Velocity magnitude \\
ANN & Artificial neural network 
    & $\mathcal{M}$ & Operator / mapping function\\
$\mathbf{W},\mathbf{b}$ & ANN weights and biases 
    & $\boldsymbol{\Theta}$ & All trainable parameters (network) \\
$\mathbf{Br}$ & Branch network 
    & $\mathbf{Tr}$ & Trunk network \\
$\zeta$ & Activation function 
    & $m$ & Total number of network layers \\
$N_l^e$ & Number of encoding layers  
    & $N_l^f$ & Number of fully-connected layers \\
$N_L$ & Number of cases in the NN input
    & $N_c$ & Number of PM classes in the NN input \\
$N_t$ & number of time points in the NN input  
    & $N_s$ & number of spatial points in the NN input \\
$N_L^{b}$ & Number of mini-batch cases 
    & $N_t^{b}$ & Number of mini-batch time points \\
$N_c^{b}$ & number of mini-batch PM classes 
    & $N_s^{b}$ & number of mini-batch sampling coordinates  \\
$N_h$ & Number of hidden neurons per layer  
    & $lr$ & Learning rate \\
$\odot$ & Hadamard (element-wise) product
    & $\circ$ & Composition of functions \\
$\mathrm{MSE}$ & Standardized mean sq.\ error  
    & $\mathrm{MSE}^\ast$ & Unstandardized mean sq.\ error \\
$\mathrm{RMSE}$ & Standardized root mean sq.\ error  
    & $\mathrm{RMSE}^\ast$ & Unstandardized root mean sq.\ error \\
$R^2$ & Coefficient of determination
    & $\gamma$ & Decay rate for learning rate \\  
$\mathcal{L}$ & Training loss
    & $\mathcal{L}_v$ &  validation loss \\

\bottomrule
\end{tabular}

\newpage
\bibliography{main.bib}

\end{document}